\documentclass[letterpaper,aps,prl,twocolumn,showpacs]{revtex4}
\usepackage{graphicx,bm}
\usepackage{amsmath,amssymb}

\begin{document}

\title{Hierarchy of relevant couplings in perturbative renormalization group transformations}
\author{Hong-Yan Shih$^1$, Wen-Min Huang$^1$, Sze-Bi Hsu$^2$ and Hsiu-Hau Lin$^{1,3}$}
\affiliation{$^1$Department of Physics, National Tsing Hua University, Hsinchu 30013, Taiwan\\
$^2$Department of Mathematics, National Tsing Hua University, Hsinchu 30013, Taiwan\\
$^3$Physics Division, National Center for Theoretical Sciences, Hsinchu 30013, Taiwan}
\date{\today}

\begin{abstract}
The phase diagram for the interacting fermions in weak coupling is described by the perturbative renormalization group equations. Due to the lack of analytic solutions for these coupled non-linear differential equations, it is rather subtle to tell which couplings are relevant or irrelevant. We propose a powerful classification scheme to build up the hierarchy of the relevant couplings by a scaling Ansatz found numerically. To demonstrate its superiority over the conventional classification for the relevant couplings, we apply this scheme to a controversial phase transition in the two-leg ladder and show that it should be a non-trivial crossover instead. The scaling Ansatz we propose here can classify the relevant couplings in hierarchical order without any ambiguity and can improve significantly how we interpret the numerical outcomes in general renormalization group methods.
\end{abstract}

\pacs{71.10.Fd, 71.10.Hf, 71.27.+a, 71.10.Pm}

%71.10.Fd	Lattice fermion models (Hubbard model, etc.)

%71.10.Hf Non-Fermi-liquid ground states, electron phase diagrams and phase transitions in model systems

%71.27.+a	Strongly correlated electron systems; heavy fermions

%71.10.Pm Fermions in reduced dimensions (anyons, composite fermions, Luttinger liquid, etc.) (for anyon mechanism in superconductors, see 74.20.Mn)

%74.20.Mn Nonconventional mechanisms (spin fluctuations, polarons and bipolarons, resonating valence bond model, anyon mechanism, marginal Fermi liquid, Luttinger liquid, etc.)

%74.20.Mn	 Nonconventional mechanisms (spin fluctuations, polarons and bipolarons, resonating valence bond model, anyon mechanism, marginal Fermi liquid, Luttinger liquid, etc.)

\maketitle

Renormalization group\cite{Wilson83,Shankar94,Fisher98,Goldenfeld92} is a powerful method to determine the effective interactions for the complex system in low-energy limit.
By integrating out the degrees of freedom at the longer length scale, the couplings describing the effective interactions flow according to a set of RG equations.
This approach has been successfully applied to a wide variety of physical phenomena, including the transport theory in the presence of impurity scattering\cite{Aleiner06,Liu09}, ground-state properties for cold atoms\cite{Mathey06,Gubbels08}, phase diagrams for iron pnictides\cite{Wang09}, Kondo lattice\cite{Ong09} and other correlated systems.
RG is particularly helpful when the quantum fluctuations in the system are strong\cite{Fabrizio93,Balents96,Schulz96,Arrigoni96,Lin97,Lin98a,Szirmai06,Lin08} so that the mean-field description is invalid or hard to justify. For instance, it has been demonstrated that, despite of the repulsive interactions at the short length scale, electrons form pairs with unconventional $d$-wave symmetry in the low-energy limit in the two-leg ladder\cite{Balents96,Lin98a}. In addition, the RG analysis also predicts that the carbon nanotubes are Mott insulating spin liquid\cite{Balents97,Krotov97,Lin98b,Odintsov99,Nersesyan03}, which is beautifully realized in recent experiments\cite{Deshpande09}.

\begin{figure}
\centering
\includegraphics[width=8.5cm]{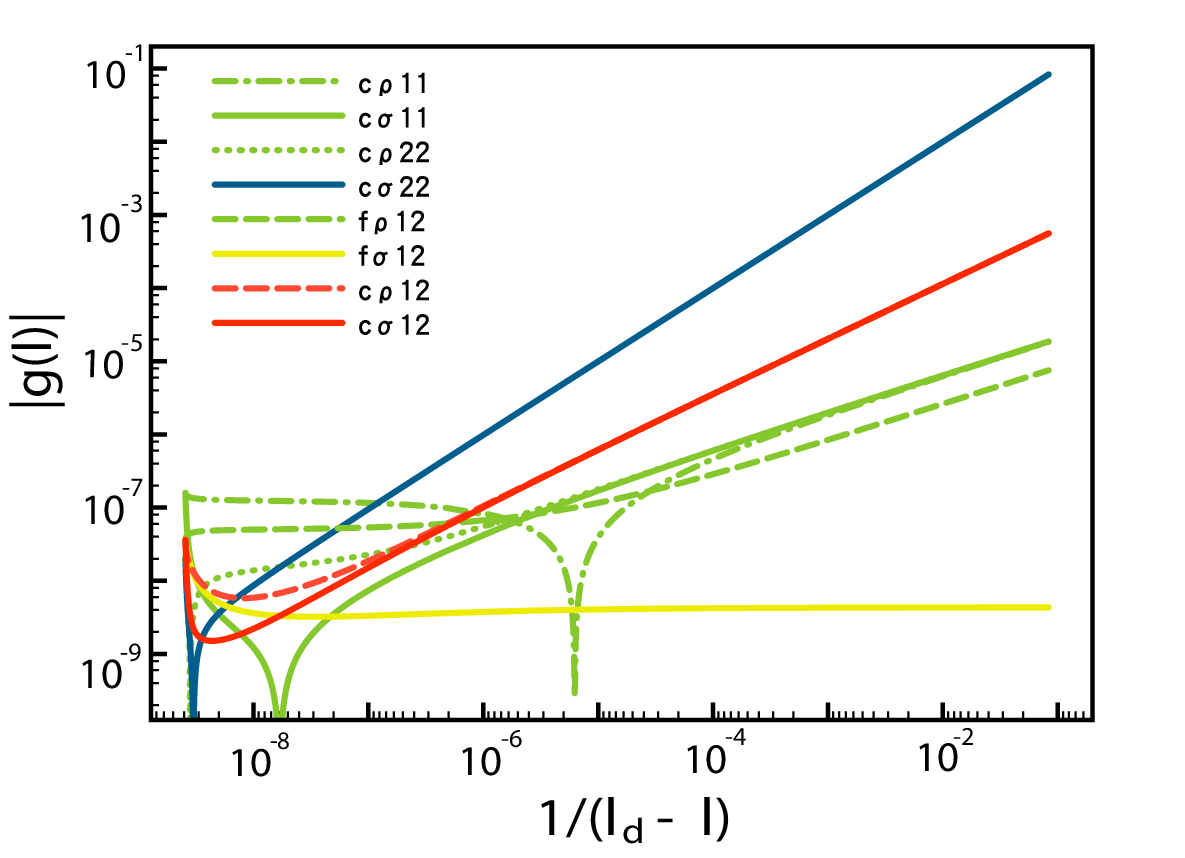}
\caption{RG flows of the eight couplings in the doped two-leg ladder. The initial couplings are generated from the on-site interaction $U/t=10^{-5}$ and the velocity ratio $v_2/v_1 = 7.8$.
We only plot the flows within the perturbative regime where all couplings $g_i < 1$.}\label{Ansatz}
\end{figure}

In weak coupling, the phase diagram for the ground states of the interacting fermions is described by the coupled non-linear RG equations,
\begin{eqnarray}
\frac{d g_i}{dl} = \Delta_{ij} g_j + A_i^{jk} g_j g_k + B_i^{jkl}g_j g_k g_l + ...,
\end{eqnarray}
where $g_i$ are the couplings for the effective interactions and $l$ is the logarithmic length scale.
$\Delta_{ij}$ denotes the scaling dimensions at the tree level and $A^i_{jk}$, $B^i_{jkl}$ are the one-loop and the two-loop renormalization.
While the derivation of the RG equations are rather standard and can be found in the literature\cite{Shankar94,Honerkamp01,Lin05}, interpreting the RG flows obtained in numerics can be subtle and tricky.
When the tree-level contributions are not zero, reading off the relevant couplings are rather straightforward\cite{Shelton96,Tsuchiizu02}.
The challenge arises when all couplings become marginal, i.e. $\Delta_{ij}=0$.
The standard recipe is to integrate the RG equations up to the cutoff length scale $l_c$ where the maximal coupling is of order one.
At $l = l_c$, we identify the couplings $g_i(l_c) \sim {\cal O}(1)$ to be relevant while those couplings $g_i(l_c) \ll 1$ as irrelevant.
But, this face-value classification scheme at the cutoff length scale does not always work and some couplings are ambiguous to be identified as either relevant or irrelevant.

To amend this ambiguity, Ledermann, Le Hur, and Rice\cite{Rice00} come up with a clever method to patch the RG flows at different length scales and obtain a hierarchy of gap opening due to different Fermi velocities.
Inspired by their success in building up the hierarchy of relevance in the RG flows, we analyze the RG equations for correlated electron systems in weak coupling numerically.
After extensive numerical studies, we find that the relevant couplings can be well captured by the scaling Ansatz\cite{Arrigoni96}
\begin{eqnarray}
g_i(l) \approx \frac{G_i}{(l_d - l)^{\gamma_i}},
\label{eq:Ansatz}
\end{eqnarray}
where $G_i$ are some non-universal constants and $l_d$ is the divergent length scale from the one-loop RG equations.
The relevant couplings are thus captured by the exponents $0 \leq \gamma_i \leq 1$ appearing in the scaling Ansatz as shown in Fig. 1.
It may sound odd at first glance why the divergent length scale $l_d$ appears in the scaling Ansatz for RG flows in the {\em perturbative} regime. This is due to an approximate scaling relation in weak coupling which we will come back later.

To demonstrate the validity of the scaling Ansatz, we revisit the phase diagram of the doped two-leg ladder.
In previous study\cite{Balents96}, it was shown that the different Fermi velocities can lead to a quantum phase transition from the spin liquid to another gapless phase.
However, in later numerical studies\cite{Noack96,Daul98}, there is no hint for the phase transition.
Therefore, it remains controversial whether the difference in Fermi velocities can drive a quantum phase transition.
It was shown before that there are eight independent couplings, $\bm{g} = (c^\rho_{11}, c^\sigma_{11}, c^\rho_{22}, c^\sigma_{22}, c^\rho_{12}, c^\sigma_{12}, f^\rho_{12}, f^\sigma_{12})$, describing the forward and Cooper scattering in the system.
Readers who are interested in details are encouraged to read the previous publications\cite{Balents96,Lin05}.

\begin{widetext}
After coarse-graining the fluctuations at shorter length scale, the one-loop RG equations for the couplings are
\begin{eqnarray}
\frac{d c_{ij}^{\rho}}{dl}&=&-\sum_{k}\frac{\alpha_{ijk}}{4}\left\{c_{ik}^{\rho}c_{kj}^{\rho}+3c_{ik}^{\sigma}c_{kj}^{\sigma}\right\}+\frac{1}{4}\left(c_{ij}^{\rho}h_{ij}^{\rho}+3c_{ij}^{\sigma}h_{ij}^{\sigma}\right),\\
\frac{dc_{ij}^{\sigma}}{dl}&=&-\sum_{k}\frac{\alpha_{ijk}}{4}\left\{c_{ik}^{\rho}c_{kj}^{\sigma}+c_{ik}^{\sigma}c_{kj}^{\rho}+2c_{ik}^{\sigma}c_{kj}^{\sigma}\right\}+\frac{1}{4}\left(c_{ij}^{\rho}h_{ij}^{\sigma}+c_{ij}^{\sigma}h_{ij}^{\rho}-2c_{ij}^{\sigma}h_{ij}^{\sigma}\right),\\
\frac{df_{ij}^{\rho}}{dl}&=&\frac{1}{4}\left[(c_{ij}^{\rho})^2+3(c_{ij}^{\sigma})^2\right],\\
\frac{df_{ij}^{\sigma}}{dl}&=&-(f_{ij}^{\sigma})^2+\frac{1}{2}\left[c_{ij}^{\rho}c_{ij}^{\sigma}-(c_{ij}^{\sigma})^2\right],
\end{eqnarray}
where $v_i$ are the Fermi velocities and the short-hand notations are defined, $h_{ij}\equiv2f_{ij}+\delta_{ij}c_{ii}$ and $\alpha_{ijk}=(v_i+v_k)(v_k+v_j)/2v_k(v_i+v_j)$.

\begin{figure}
\centering
\includegraphics[width=17cm]{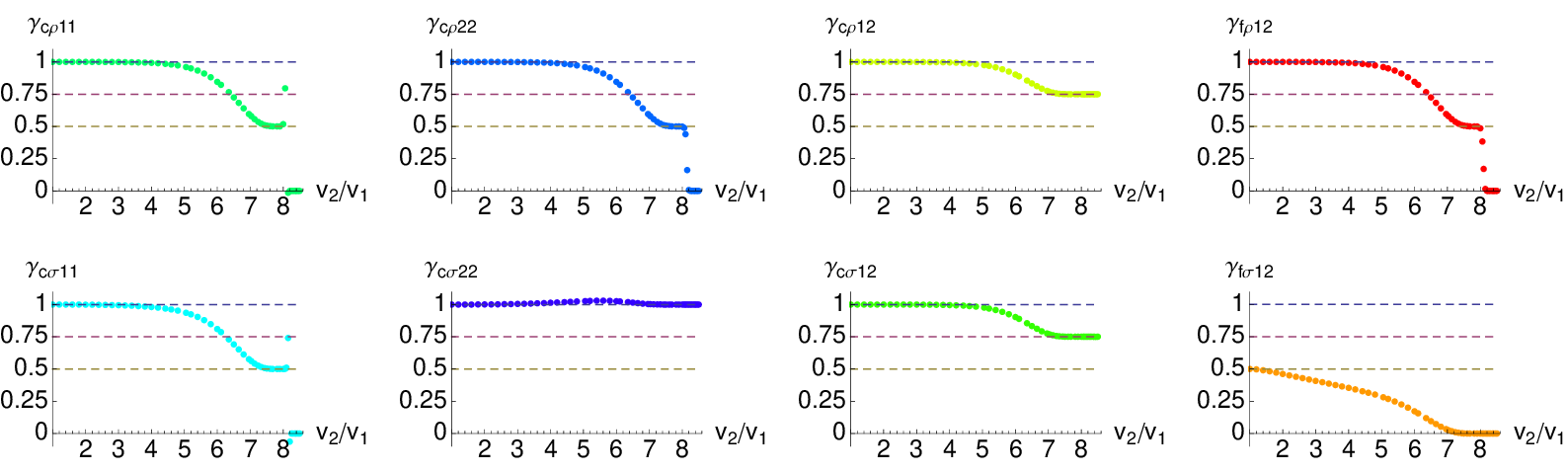}
\caption{Exponents extracted from the numerical solutions for the RG equations at different velocity ratios $v_2/v_1$ in the scaling regime. For small velocity ratios, all couplings except $f^\sigma_{12}$ are equally dominant with exponents $\gamma_i=1$. As the velocity ratio increases, the relevance of the couplings can be classified by the exponents accordingly.}\label{Exponents}
\end{figure}
\end{widetext}

A typical example for the RG flows with the on-site interaction $U/t=10^{-5}$ and the velocity ratio $v_2/v_1 = 7.8$ is shown in Fig.~\ref{Ansatz}.
First of all, we can extract the divergent length scale $l_d \sim 10^7$ numerically.
At the beginning, the RG flows are rather messy and it is hard to tell which are the relevant couplings.
However, as the short-range fluctuations are integrated out gradually, the couplings enter the scaling regime described by Eq.(\ref{eq:Ansatz}) with different exponents $\gamma_i$.
The straight lines in Fig.~\ref{Ansatz} indicate that the scaling Ansatz works extremely well.
It is worth emphasizing that all renormalized couplings remain in the perturbative regime even though the Ansatz mysteriously contains the divergent length scale $l_d$.

The exponents provide an alternative way to classify the relevant couplings.
The eight couplings fall into four categories: (1) the predominant group $\gamma_{c^\sigma_{22}}=1$, (2) the first subdominant group $\gamma_{c^\rho_{12}} = \gamma_{c^\sigma_{12}}=3/4$, (3) the second subdominant group $\gamma_{c^\rho_{11}}=\gamma_{c^\sigma_{11}}= \gamma_{c^\rho_{22}} = \gamma_{f^\rho_{12}}=1/2$, (4) the subsidiary group $\gamma_{f^\sigma_{12}}=0$.
By extracting the exponents from the trends of the RG flows, we can build up the hierarchy of relevance without any ambiguity.

The classification scheme can be applied to different velocity ratios with exponents summarized in Fig.~\ref{Exponents}.
Upon doping away from the half filling, the velocity ratio $v_2/v_1$ starts from unity and gradually increases.
When the velocity ratio is close to one, all exponents $\gamma_{i}=1$ except that for $f^\sigma_{12}$.
It is not surprising that the conventional face-value classification scheme works rather well in this regime since all couplings are equally relevant.
Applying the bosonization technique, the interactions related to gap opening are
\begin{eqnarray}\label{bosonization}
&&\nonumber\hspace{-0.9cm}{\cal H}_I\sim\hspace{0.1cm}\hspace{-0.05cm}c_{11}^{\sigma}\cos\left(\sqrt{2}\theta_{1\sigma}\right)+c_{22}^{\sigma}\cos\left(\sqrt{2}\theta_{2\sigma}\right)\\ &&\hspace{-0.2cm}+\hspace{0.1cm}4c_{12}^{\sigma}\cos\left(\frac{\varphi_{1\rho}-\varphi_{2\rho}}{\sqrt{2}}\right)\cos\left(\frac{\theta_{1\sigma}}{\sqrt{2}}\right)\cos\left(\frac{\theta_{2\sigma}}{\sqrt{2}}\right).
\end{eqnarray}
Here we express the Dirac fermion fields in term of chiral boson fields, $\psi_{Pi\alpha}=\kappa_{i\alpha}e^{i\phi_{Pi\alpha}}$ with the Klein factors $\kappa_{i\alpha}$\cite{Lin98a}.
Meanwhile, by introducing the conjugate boson fields for each flavor, $\varphi_{i\alpha}=\phi_{Ri\alpha}+\phi_{Li\alpha}$ and $\theta_{i\alpha}=\phi_{Ri\alpha}-\phi_{Li\alpha}$, we separate the boson field into charge and spin modes, $\theta_{i\rho}=\left(\theta_{i\uparrow}+\theta_{i\downarrow}\right)/\sqrt{2}$ and $\theta_{i\sigma}=\left(\theta_{i\uparrow}-\theta_{i\downarrow}\right)/\sqrt{2}$ (similar for the conjugate field $\varphi$).
Since the couplings $c^\sigma_{11}, c^\sigma_{22}, c^\sigma_{12}$ are relevant, spin gaps $\Delta_1$ and $\Delta_2$ for the antibonding and bonding bands develop under RG transformation.
The magnitudes of the spin gaps can be estimated by the semiclassical approximation, $\Delta_1/\Delta_2\sim\left(|c_{11}^\sigma|+c_{12}^\sigma\right)/\left(|c_{22}^\sigma|+c_{12}^\sigma\right)$, where the values of the couplings are taken at the cutoff length scale.
The detail bosonization analysis shows that the ground state is a spin liquid with unconventional $d$-wave pairing between electrons\cite{Balents96,Lin97,Lin98a}.

As the velocity ratio increases $v_2/v_1 \gtrsim 5$, $\gamma_{c^\sigma_{22}}$ remains one but the other exponents start to fall, as shown in Fig.~\ref{Exponents}.
The four-level hierarchy of relevance described before starts to emerge.
When the velocity ratio further increases $v_2/v_1 \gtrsim 8$, the second subdominant exponents also falls to zero, $\gamma_{c^\rho_{11}}=\gamma_{c^\sigma_{11}}= \gamma_{c^\rho_{22}} = \gamma_{f^\rho_{12}}=0$ but the predominant and the first subdominant exponents remain intact.
The exponents of the scaling Ansatz provide a simple and elegant way to classify the relevant couplings.
It is also interesting that the evolution of the exponents mainly stays on plateaus of simple ratio numbers.
These ratio numbers can be obtained by performing linear stability analysis\cite{Lin05} near the fixed rays of the RG flows.
However, since the initial couplings are nowhere close to the fixed rays, it is not clear why the RG flows inherit the simple rational exponents.
The plateau structure of the exponents seen in numerics remain open for further investigations.

Within our classification scheme, we find the pattern for the excitation gaps is the same except the magnitude of the spin gap in the bonding band $\Delta_2$ is much larger than $\Delta_1$ in the antibonding band.
Therefore, as the velocity ratio $v_2/v_1$ increases, the gap ratio $\Delta_2/\Delta_1$ also increases significantly but there is no phase transition for $1 < v_2/v_1 < 8.6$.
However, the conventional read-off scheme at the cutoff length scale misses all the subdominant couplings and misidentify the crossover as a quantum phase transition\cite{Balents96,Lin97}.
In addition, the scheme proposed here naturally leads to a hierarchy of excitation gaps predicted by the patched RG method\cite{Rice00}.

\begin{figure}
\centering
\includegraphics[width=8cm]{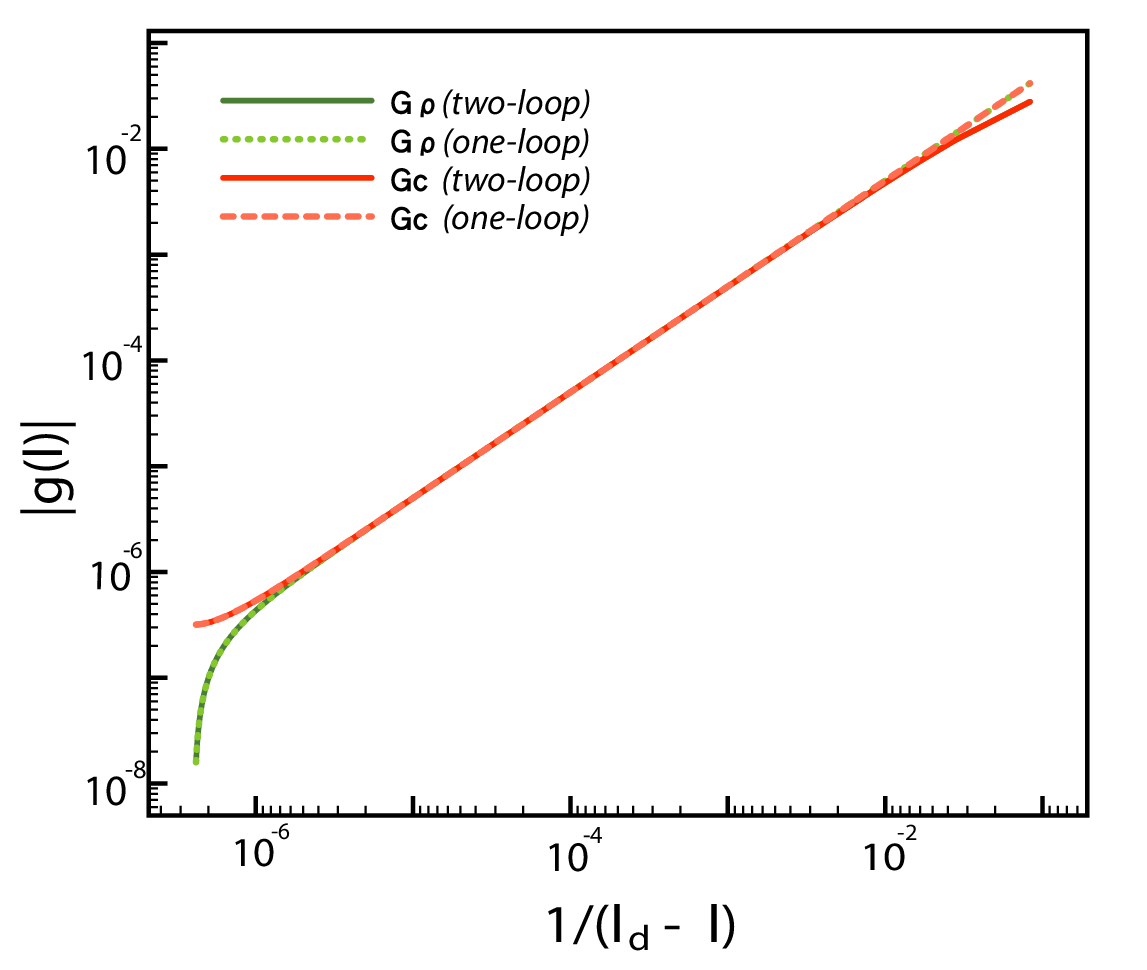}
\caption{RG flows up to two-loop order for the charge sector in the one dimensional Hubbard model.}
\label{two-loop}
\end{figure}

We also try out the scaling Ansatz beyond the one-loop order and found it remains valid. For instance, the RG equations for the one dimensional Hubbard model\cite{Tsuchiizu06} are
\begin{eqnarray}
\frac{dG_\rho}{dl} &=& 2 G_c^2 - 2 G_\rho G_c^2,
\nonumber\\
\frac{dG_c}{dl} &=& 2 G_\rho G_c -2 G_\rho^2 G_c - G_c^3.
\end{eqnarray}
Here we only put down the charge sector since the RG equations for the spin sector are decoupled with similar structure.
As is evident from Fig.~\ref{two-loop}, the two-loop corrections are negligible and both $G_\rho$ and $G_c$ follow the scaling Ansatz rather well.
Therefore, the scaling Ansatz and the corresponding exponents remain valid beyond the one-loop order as long as the flows stay in the perturbative regime.

Now we come back to the puzzle about the divergent length scale $l_d$ in the scaling Ansatz.
This arises from a non-trivial relation for the one-loop RG flows, $g_i(l) = U G_i(Ul)$, where $G_i(l)$ is the solution with order-one initial values while $g_i(l)$ is the solution in weak coupling (with initial values of order $U \ll 1$).
Thus, there is a non-trivial connection between the RG flows in perturbative regime and those in the singular regime.
This relation may seem an artifact for the one-loop RG equations.
But, it is rather remarkable that the relation remain approximately correct even when higher-loop corrections are included.
This is also consistent with the numerical findings that the RG flows are dominated by the one-loop terms with negligible corrections from the higher loops.

Now we try to see how the scaling Ansatz emerges in the non-linear RG equations.
It is insightful to rewrite the equations in matrix form $dg_i/dl = \bm{g}^{T} \bm{A}_i \bm{g}$, where the real symmetric matrices are $(\bm{A}_i)_{jk} = A_i^{jk}$.
Assume that the matrices $A_i$, $i=1,2,..,n$, are positive definite.
One can use the standard decomposition to rewrite the matrix $\bm{A}_n = \bm{L}_n \bm{D}_n \bm{L}_n^{T}$, where $\bm{L}_n$ is a unit $n\times n$ lower-triangular matrix and $\bm{D}_n$ is a diagonal Gaussian pivot.
Simple algebra leads to the important inequality,
\begin{eqnarray}\label{gn}
\frac{dg_n}{dl} \geq \Delta_n g_n^2,
\end{eqnarray}
where $\Delta_n > 0$ is the $n$-th pivot in the diagonal matrix $\bm{D}_n$.
Integrating the inequality in Eq.~(\ref{gn}), one can show that the RG flows become singular at some finite length scale $l=l_d$.
Following similar algebra, one can also show that all couplings become singular at exactly the same length scale. 

The above proof establish the existence of the singular length scale $l_d$ in mathematical rigor, making the scaling Ansatz with power-law singularity plausible but not yet proven.
However, if one plugs in the scaling Ansatz into the inequality, it leads to the constraint $\gamma_i \leq 1$ as found in the numerics.
The situation is further complicated by the fact that the matrices $\bm{A}_i$ in the RG equations are not always positive definite.
This is reasonable because, for some initial coupling profile, the RG flows may not become singular at all.
However, as long as the initial couplings sit inside the unstable manifold in the multi-dimensional coupling space, the above statements are expected to be correct.

Since the classification scheme we proposed here is rather general, the numerical solutions from the functional RG approach\cite{Honerkamp01} can be classified by the scaling exponents as well.
In fact, we apply the scaling Ansatz to a recent RG analysis\cite{Chubukov08} for iron pnictides.
There are four couplings $(u_1, u_2, u_3, u_4)$ describing the interactions between the electron and hole pockets.
We numerically integrate these RG equations and find the scaling Ansatz works again with exponents $\gamma_i = 1,\frac13, 1,1$.

In conclusion, we propose a powerful scheme to classify all relevant couplings in the one-loop RG equations. As long as the initial couplings are weak, the renormalized couplings enter the scaling regime characterized by a unique set of exponents $\gamma_i$. We can build up a hierarchy of relevance by these exponents without ambiguity and greatly improve the interpretations of the results obtained by various RG methods.

We acknowledge supports from the National Science Council in Taiwan through grant NSC-97-2112-M-007-022-MY3. Financial supports and friendly environment provided by the National Center for Theoretical Sciences in Taiwan are also greatly appreciated.

\end{document}